# Sustaining Knowledge Infrastructures: Asking the Right Questions and Listening for Answers

Kathleen Gregory, Centre for Science and Technology Studies, Leiden University
Jonathan Zurbach, Avignon University
Kalpana Shankar, School of Information and Communication Studies, University College Dublin
Matthew Mayernik, NSF National Center for Atmospheric Research
Malcolm Campbell-Verduyn, University of Groningen
Louise Bezuidenhout, Centre for Science and Technology Studies, Leiden University
Andrew Treloar, Australian Research Data Commons

## Abstract

Sustaining knowledge infrastructures (KIs) remains a persistent issue that requires continued engagement from diverse stakeholders as new questions and values arise in relation to KI maintenance. We draw on existing academic literature, practical experience with KI projects, and our discussions at a 2024 workshop for researchers and practitioners exploring KI evaluation to pose five questions for KI project managers to consider when thinking about how to make their KIs evolve sustainably over time. These questions include reflecting on sustainability throughout the life cycle of KIs, communicating evolving visions and values, engaging communities, "right sizing" a KI, and developing an iterative process for decision-making. Reflecting on these themes, we suggest, can support KI stakeholders to evolve (not necessarily "grow") to meet the needs and values of their communities. How these themes are discussed will necessarily vary by funding sources, discipline(s), governance, communities, and other contextual factors. However, adopting a deliberate and strategic approach to KI sustainability and aligning the invisible infrastructural work of KI maintenance with the outward-facing institutional work is, we argue, relevant to all KIs

Correspondence should be addressed to Author Name, Postal Address. Email: email@ddress

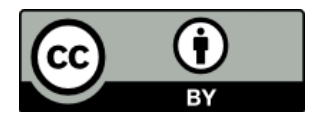

# Introduction

In their introduction to a special issue of the Journal of the Association for Information Systems on infrastructure studies, Edwards, et al. (2009) wrote that

> 'Infrastructure projects dominate economic stimulus proposals; repairs to aging bridges and roads compete intensively with investments in renewable energy sources and electronic medical records. Infrastructure today seems both an all-encompassing solution and an omnipresent problem, indispensable yet unsatisfactory, always already there yet always an unfinished work in progress.'

Sixteen years since that writing, these issues of infrastructural sustainability are still very much with us. This is the case not just for physical infrastructures, but increasingly for the knowledge infrastructures (KIs) that underpin the institutions serving research and policy bases. Such infrastructures are much more than technological systems; rather, KIs are complex networks of people, content, technology, and institutions supporting knowledge production (Edwards, 2010).

KIs were once built – and their ongoing maintenance justified – primarily for their ability to support science and innovation. Now, managers, funders, and users are asking different questions about KI maintenance and the values underpinning them. Such questions are driven by policy shifts, technological and institutional changes, and new collaboration and governance models. These increasingly emphasize the need to consider openness, ethics, inclusivity, and sustainability.

In April 2024, a group of twenty-two knowledge infrastructure researchers and practitioners came together in Leiden, the Netherlands for a four-and-a-half-day workshop funded by the Lorentz Center and the Institute for Advanced Studies at the University of Amsterdam[1]. The workshop built on and expanded work done in two previous Knowledge Infrastructure (KI) workshops, held in 2012 and 2020. At the 2024 workshop, we collectively explored the following question through a mixture of presentations, case studies, and group work: How can we evaluate knowledge infrastructures (KIs) in the context of emerging issues, such as sustainability, open science, citizen science, and threats to democracy?

In this piece, we focus on one particular dimension of the evaluation of KIs: sustainability. We acknowledge that there are multiple ways to think about sustainability: the medium and long-term preservation of KIs, the contributions of particular KIs to addressing global climate change, and the interplay of the technical, financial and human resources that go into maintaining and using KIs. We therefore conceptualize sustainability as a process of evolution, often undertaken as a KI works towards self-preservation, which is influenced by the complex network of relations in which a KI is embedded. We argue that thinking about sustaining a knowledge infrastructure requires ongoing, critical attention that can be framed by thinking about key questions, particularly as a KI transitions from being a project toward becoming an organization. We further argue that sustaining a KI requires a mix of both day-to-day infrastructural work and institutional work with established and influential actors (Mayernik, 2023).

There are numerous existing reports on how to sustain KIs and models for how to do this effectively (Directorate-General for Research and Innovation, 2017; ERIC Forum, 2022; ESFRI, 2017; OECD, 2017). The notion of "lifecycle" is commonly used in these reports and models to explain how stakeholders of a KI may work to ensure its sustainability by transitioning a KI from the state of a project to that of an organization or institution. This reflects the dynamic nature of the milieu in which KIs are embedded and is helpful in understanding KI sustainability as a process of institutional stabilization mobilizing both organizational compliance and innovation. It could be argued that the lifecycle perspective lacks a "humanistic

1 More details about the aims, outcomes, and participants at this workshop can be found in the full workshop report (Mayernik et al., 2025)

approach" (Verhulst, 2024, p. 98) that takes into account the situation of KI managers as individuals, who may struggle with the meaning of KI sustainability and question its relevance as a guide for decision-making about the future.

Following Verhulst's (2024) invitation to understand questions not merely as inquiry tools but rather as "governance devices" enabling a "polycentric perspective" on digital infrastructures, we pose one pre-question and five questions for KI managers to consider regarding sustainability. In line with Verhulst (2024), we contend that governing a KI needs to take into account more than funders' administrative requirements or the needs of a community. The governance needed to sustain a KI also requires KI managers to have the competence, agency, and judgment to ask the right questions at the right time.

Our questions start from the perspective of addressing sustainability during the transition of a KI from a project to a more institutionalized infrastructure. We seek to cultivate an awareness of how sustainability manifests as a challenge at different timepoints in this evolution. Through the process of question-making, our contribution aims to show that individual agency and interpretation matter as much as broader organizational dynamics. We also highlight how 'sustainable evolution'  requires multiple types of work, including the institutional work of formulating (and reformulating) policies, plans, workflows, and responsibilities, along with maintenance work in an infrastructure.

In this reappropriation of the sustainability question by KI managers, the cardinal question is the ''why": Why does sustainability matter for your KI? (Why) is it important that your infrastructure project withstands the test of time? How will the answer to this question itself change over time? Before addressing this broader issue, which encompasses our five questions, we begin by clarifying what we mean by knowledge infrastructures.

# What are Knowledge Infrastructures? (Our 'Pre Question')

For the purposes of this article, we define knowledge infrastructures (KIs) as "robust networks of people, artifacts, and institutions that routinely generate, share, and maintain specific knowledge about the human and natural worlds" (Edwards, 2010, p.17). The original definition was directed at infrastructures designed to routinely produce specific knowledge, such as weather forecasts, labor statistics, or pandemic tracking. Previous work distinguished systems (centrally designed and controlled) from networks (similar systems, interconnected according to a common standard or by a central agent) and webs or internetworks, formed by disparate agents by joining resources belonging to different systems or networks, with little or no centralized authority or control (Edwards et al., 2007).

In the 2024 Lorentz workshop, we did not strictly maintain these distinctions, in part because they are not always recognised or identified by the infrastructures themselves. Instead, and in line with our workshop aims, we discussed a wide range of both established and emergent systems and networks that facilitate knowledge production. Data or information infrastructures, such as repositories or libraries, support research work broadly and have been established for such purposes for decades, if not centuries. Some research infrastructures are more focused on supporting particular disciplines or areas of inquiry. Some emerging infrastructures are best characterized as platforms that present multiple pre-existing systems and data resources (e.g. demographics, surveys, and statistics generated by government agencies) within a common framework, aiming to make it easier for researchers to find and use any or all of them. The Dutch Open Data Infrastructure for Social Science and Economic Innovations[2] (ODISSEI, presented at the workshop) is an excellent example of a research infrastructure that also positions itself as a data platform. We also noted in the workshop that KIs had moved beyond

2 https://odissei-data.nl/

their initial focus on just facilitating the work of specific research disciplines to also facilitate knowledge sharing among researchers, citizens, policymakers, and others.

Here, we draw on examples of different types of KIs presented and discussed at the Lorentz workshop and the rich practical and conceptual literature exploring the sustainability of research and knowledge infrastructures, to propose five questions for KI managers to consider regarding the institutional work required to address sustainability challenges.

# 1. When Should You Talk About Sustainability?

The short answer is, somewhat unhelpfully: throughout the life cycle of the KI, from project to organization. However, concerns and questions about infrastructural sustainability arise most abruptly when managers of emerging infrastructure projects begin to consider the conditions necessary to ensure that their projects withstand the test of time in the long run (or if they should).

From this temporal dimension stems the conceptualization of KIs as dynamic and evolving, rather than as static entities. When strategizing for how KIs can evolve sustainably, KI managers should consider: How can people and work practices be arranged in the longer term? How can a KI strategically plan for and evaluate sustainability? Should (and if so how, and when) a KI project transition to becoming an organization through a process of institutionalization?

From the 19th century onwards, as Edwards points out, stakeholders of large-scale infrastructure projects designed to generate global information have had to ensure these projects are enduring in order to make them successful (Edwards, 2010). According to Edwards, the endurance of an infrastructure has intrinsic value because it "legitimises the knowledge it produces" and contributes to transforming the infrastructure into a self-perpetuating entity (Edwards, 2010, p. 25). Thus, sustainability is embedded in the very definition of what qualifies as a KI, since to exist as a KI, an entity must be, first and foremost, "robust" (Edwards, 2010, p. 17) , meaning it is capable of withstanding the test of time by evolving to be anchored in functions relevant to knowledge practices.

In their reflections on the meaning of "organisational sustainability" in the field of digital cultural heritage, Eschenfelder et al. suggest that key elements to consider are "the arrangements of people and work practices that keep digital projects and services going over time, given ongoing challenges" (Eschenfelder et al., 2018, p. 183). Regarding the management dimension of organisational sustainability, the most prominent aspect is strategic planning, which happens at regular timepoints "well beyond the initial phases of a project" (Eschenfelder et al., 2018, p. 188). Once again, project sustainability is seen as a "positive thing", and the closures of projects are framed as something to be avoided if at all possible (Eschenfelder et al., 2018, p. 193). What these examples illustrate is not surprising: sustaining a KI is a series of intentional activities of planning, implementation, and evaluation; in other words, a process of institutional work conducted by individuals asking the right questions.

To give another example, the EarthCube program, funded by the US National Science Foundation from 2011 to 2023, commissioned a study of existing data infrastructures, to see how they had (or had not) addressed important sustainability questions (Virapongse et al., 2024). In their study of long-term data infrastructure projects in earth sciences, the authors identified a pattern: for an infrastructure *project* to sustain its value beyond its initial funding period, it must evolve into an *organization* — "a legal and administrative body that separates individual goals from the goals of the project itself" (Virapongse et al., 2024, p.2). Infrastructural sustainability, demonstrated by longevity, is thus defined by the authors as "the ability of a project to maintain its core mission and functions beyond the initial seed or start-up funding, and despite a change over time in the individuals who make up the organization" (Virapongse et al., 2024, p.2). Transforming an infrastructure project into an organization enables it to withstand the test of time, by forcing the people involved to create formal structure around business models, leadership and professional positions, and community engagement.

These three examples illustrate that exploring KI sustainability requires moving away from viewing infrastructures as "predominantly inert(ial) objects" and compels us to adopt a longitudinal viewpoint that incorporates the sociotechnical systems in which KIs are embedded as well as the individuals working to shoulder KIs (Mounier & Dumas-Primbault, 2023, p. 12). Examining the sustainability of KIs, especially as projects, necessitates considering their "persistence as infrastructures" (Mounier & Dumas-Primbault, 2023, p. 24). Thus, asking and answering sustainability questions is a perpetually unresolved and renewed issue: each time a new data, information, or knowledge infrastructure project emerges or is proposed, its long-term sustainability will inevitably be called into question, and a response will need to be formulated.

For a KI project to transition into an organization—that is, to increase its agency—questions of sustainability relevant to the project as such must be addressed. As suggested by Virapongse et al., the individuals who formulate sustainability questions should provide sufficiently robust organizational responses to enable them to pass responsibility on to others, who will in turn pose new sustainability questions without endangering the continuity of the KI's functions. When governing a KI project, asking sustainability questions thus appears as a means of ensuring a degree of independence between those who formulate the questions and the future of the KI. The increase in agency of a KI resulting from institutional work therefore entails the collaboration of successive cohorts of KI managers, each asking the right questions at different stages of the KI lifecycle and thereby building a legacy of sustainability-oriented responses.

As long as stakeholders involved throughout the process of infrastructural evolution continue to champion infrastructure projects, the question of their sustainability will remain relevant. KI sustainability, especially on a large scale, cannot be achieved overnight. Instead, it requires a long, sustained effort and is an arduous task aimed at fostering endurance, robustness, longevity, and persistence. In other words, making a project dedicated to scientific knowledge or research sustainable involves intentional individual and collective work focused on ensuring it can endure over time. However, intentionality in institutional work also requires the ability to listen to answers to sustainability questions relevant to KI formulated by other actors beyond the purview of our workshop, whether these answers concern legal structures, funding models or technological choices. Consequently, KI project managers, in their struggle with the meaning of sustainability questions, as evidenced during the discussion at our workshop, cannot address sustainability questions alone; these issues typically need to be discussed with project funders and policymakers.

# 2. How Can you Communicate the Vision and Value of your KI?

Effectively communicating the vision of a KI to different stakeholders, particularly to funders and policymakers, is a key part of moving from an infrastructure project to becoming a more sustainable organization. This involves engaging in both "infrastructural work" related to day-to-day operations and "institutional work" with established actors who have power and influence in the development of KIs (Mayernik, 2023). It also requires learning how to speak a common language when pitching the vision for the future of a KI. From the standpoint of institutions, such as funding or policymaking organizations, this often boils down to communicating a financial vision and demonstrating long term "value."

Key questions to consider for KI managers along this line include: How will you secure and maintain funding? How will you communicate strategic (financial) plans to stakeholders? How will you weave such considerations into daily tasks and routines? How can you come together with other stakeholders to engage in "institutional work" to strengthen relations and communicate the vision of your KI?

Since the 1950s, ensuring the longevity of large-scale scientific infrastructure projects – and making the case for their value to key stakeholders who hold the purse strings – has been a key concern for project managers. From a strictly economic perspective, the decision to make a KI sustainable can be understood in terms of a cost-benefit analysis. Such an analysis is generally conducted by a public or private funder based on the relevance of a value proposition made by the project's stakeholders regarding goals defined by science policymakers (Florio, 2019). Even if the financial viability of a large project in the long term is not the only issue addressed by managers in discussions with funders or policymakers about how to transition from emerging infrastructure projects to sustainable infrastructures (Chan, 2019), all other matters hinge on the answer to this primary question. As Jacob & Hallonsten (2012) demonstrate, the very existence of "megascientific" infrastructure projects in the long run depends on the persistence of investments in such projects (though financial resources alone are necessary but not sufficient).

Maintaining investments in KIs requires that managers demonstrate the long-term value of KIs to other stakeholders; this has now become an expectation or even mandatory requirement from science funders and policymakers. As Mounier & Dumas-Primbault (2023) suggest, a crucial logic influencing — or even pressuring — the governance of KIs stems from legitimizing discourses required by funders and policymakers to justify both the funds allocated to projects and the budgets for infrastructure continuity. To secure funding, KI managers must therefore explicitly address issues of sustainability — KI funders now expect sustainability plans. Having such a plan is in fact part of the process of demonstrating the legitimacy of a KI.

As a whole, this means that KI managers need to actively and continually evidence the long-term value of their projects, which may encompass financial, scientific, educational, public, or other types of value (or values, which may be less measurable). Many infrastructure managers have been assigning persistent identifiers (PIDs) as a mechanism to support measurements of impact and value (Klump & Huber, 2017). PIDs are now being assigned to datasets, software packages, research instrumentation, physical samples, high-performance computing systems, research facilities, and many other types of research entities, as well as to people and organizations who use and provide these entities (Cousijn et al, 2021). If a community does not use persistent identifiers when referencing data or an infrastructure itself, it may not be possible to develop metrics which can potentially be used to argue for the sustained need for a KI. These PID initiatives demonstrate how modes of evaluation are themselves necessary to help ensure the persistence of an infrastructure.

Ensuring and demonstrating value at its heart involves infrastructural as well as institutional work. By aligning and "coupling" iteratively high investments in infrastructural work related to "storage systems, repositories, tools, and interfaces" and institutional work related to "policies, intermediaries, governance processes, routines, and financial instruments," a "balanced and measured progress toward solutions to technical and organizational problems" can be achieved (Mayernik, 2023, p.47). The effective demonstration of KIs' long-term value (and thus sustainability) can be understood as an outcome of such a coupling between infrastructural and institutional work.

# 3. How Can You Bring Communities Together to Sustain a KI?

KI managers are of course not solely responsible for sustaining knowledge infrastructures. KI are embedded within complex networks of organizations, some of which may be permanent, others which come together to enact particular projects. Such organizational structures are complex and complicated; they co-evolve in response to internal and external pressures and needs. In turn, the communities that form within these structures can learn from each other while also acquiring feedback from other stakeholders, constituting an interorganizational community of practice (De Groot et al., 2022 ) that builds knowledge over time. Within these structures the kinds of contributions towards the KI may be funded, in-kind, or voluntary. This in turn makes it hard to quantify who is doing the sustaining, and what would be required to plan for greater sustainability.

Different (user) communities, who have diverse needs and meanings for KIs, also play a role. Sustaining an infrastructure therefore involves understanding the needs of these communities, and negotiating the different purposes which they have for KIs. Synthesizing and weaving together these meanings and needs can help KI managers to prove the value of KIs to funders and policymakers and stay relevant to different communities.

KI managers therefore need to think deeply about the imagined and actual communities for their infrastructures when considering its sustainability: Who is contributing to a KI? Who may resist or push back? Who will remain neutral? What are the actual needs of particular communities? How might these needs change and evolve?

A key component of the sustainability of a KI is its usefulness to different communities and actors (Morselli and Edmond, 2020), and its ability to anticipate unintended uses (Wofford and Thomer, 2023). Despite its importance, the notion of "communities" or "designated communities" is rarely deeply interrogated (Borgman, 2012) and is often used loosely without definition; similarly, the idea of community engagement can also be reductive, where speaking with a handful of individuals can be used as a signifier of interacting with and building on the needs of a broader community. It is also difficult to understand the actual scope of communities, where the number of "sign-ups" or website visits can lead to overestimating the size or cohesion of a community.

KIs which are created using a "build it and they will come" mentality, which have vague definitions of their communities, stand in contrast to infrastructures such as ODISSEI or DARIAH[3], which emerged from and were co-developed with communities in the social sciences and digital humanities to address specific problems. Identifying these problems required these KIs to think about and adapt the type of content which they include. ODISSEI and DARIAH now continue to engage with their evolving communities, adapting the type and format of content that is needed. They are therefore interwoven into everyday research practices, even as the infrastructures continue to grow.

Certain communities may also push back or resist a KI, which has ramifications for sustainability. One example of this is the case of a citizen science infrastructure in the Netherlands (Samen Meten) which was developed to monitor air quality using measurements collected by citizens. While many engaged citizens saw the value of this infrastructure and actively participated, others resisted it. For example, a coalition of farmers, concerned about how the monitoring data would be used to support legislation to reduce atmospheric nitrogen emission (and potentially their crop yield through constraints on fertiliser use), organised lengthy protests outside the ministry. Potentially, this contestation could have led to negative repercussions for the infrastructure in terms of securing continued funding; instead, the infrastructure became a space for continued negotiation and conversation among different community groups (Fenlon et al., 2023; Rayburn et al., 2024). Aligning infrastructural and institutional work requires therefore a balance between organizational stability and flexibility.

# 4. What is the Right Size for Your KI?

Knowledge infrastructures are often bespoke resources, designed for particular purposes, communities and content. However, despite their often very specific and targeted beginnings, it is assumed that growth of the infrastructure (in terms of an increase in content, user communities, or funds) is the ultimate goal, necessary to remain relevant and sustainable. This focus on growth can be in tension with the actual needs of various actors and the mission of a KI. For this reason, we argue that KI managers need to be mindful of whether a trajectory of growth is inevitable, or should be. They should also consider de-growth or "right-sizing" in order to meet the needs of the KI, various communities forming around the infrastructure, and the capacities of the KI itself. This also involves considering making plans for closing down a KI once it no longer serves its purpose (Bilder et al., 2015).

When considering the right size of a KI, managers should therefore ask: Is growth always the right option? What other options are there which can help to meet the aims of a KI and needs of various communities? When should a KI be sunsetted? Could parts of a KI be sunsetted usefully and successfully, while others are maintained?

3 https://www.dariah.eu/

The “right size” for a KI is of course a fluid concept that will necessarily evolve over time, and what constitutes de- or post-growth will be similarly fraught (Campbell-Verduyn and Kranke 2025). For example, social science data archives, such as ICPSR, DANS, or the UK Data Archive, have historically tended to operate separately from each other, serving individual countries or communities. The advent of standards and technologies for federation (and the need for cross-border data exchange) forced these archives to grow in size and scope (Eschenfelder et al., 2018). Other KIs have closed down as they have been superseded, defunded, or otherwise discontinued. To some extent, it is a difficult path to track the demise of a KI, as there are few documented examples of how and why KI closure has happened. Some of the ethnographic studies of decay/downsizing KIs speak to the deeply personal, socially fraught dimensions of decommissioning KIs (Jackson & Buyuktur, 2014; Cohn, 2016). As a whole, considering the right size of a KI is closely linked to understanding its relevance to particular communities and the vision of a KI itself, as well as responding to existing and envisaged resource constraints, rather than seeing expansion and growth as an unquestioned necessity.

# 5. How Can You Make Sustainability Decisions in an Ongoing, Reflective Way?

As we have seen, sustaining a KI involves individuals with extended interpretative agency, that is KI managers, making a host of decisions, e.g. about relevance, size, community engagement and finances. Some of these decisions are related to infrastructural considerations (e.g. about technical or human capacity) and others are related more to institutional work, such as decisions about governance structures, rules for participation and access, and mechanisms for user engagement. Such decisions need to be continuously revisited and re-evaluated; designing an iterative process for decision-making which includes formative points for evaluation is, we argue, another key aspect of achieving sustainability.

Key considerations for KI managers here are related to both the content of the decisions themselves as well as decision-making processes. Such questions include: Are we up to date on technical standards and community engagement methods? Do we have adequate technical and human capacity to achieve our aims? What is our governance structure (not) achieving? How can we build formative evaluations of sustainability into our daily routines? Do we need to pivot and adapt?

Existing frameworks created within research data management and scholarly infrastructure communities can be useful when thinking about the content of the decisions which need to be addressed when considering sustainability, some of which have already surfaced in our discussion thus far. One example of such a framework is the Principles for Open Scholarly Infrastructure[4] (POSI). While POSI details a number of financial considerations, it also describes points related to institutional governance which impact sustainability, e.g. a recommendation for being stakeholder-governed, to work towards transparency in institutional decisions, and to avoid lobbying for personal over communal goals.

The Global Open Research Commons Essential Elements (GORC) (Jones et al, 2023; Treloar & Woodford, 2024), a framework for the interoperability of research data infrastructures, also addresses governance as an important aspect of the sustainability of an interlinked ‘commons’ of infrastructures. The GORC International Model emphasizes that governance structures themselves must be designed in such a way to enable transition to different ways of governing as needed. This framework also identifies other ‘infrastructural’ decisions which need to be taken to ensure both sustainability and interoperability, e.g. regarding continued evaluations of data contents, ensuring adequate technical and human capacities, and the use of standards between data repositories rather than bespoke solutions.

4 https://openscholarlyinfrastructure.org/

How to iteratively make such decisions (and evaluate them over time) is arguably just as important as what to consider. Here literature about the evaluation of the 'impact' of research infrastructures over the short, medium and long term can be helpful. An approach for demonstrating the value of KIs, particularly to funders, that is increasingly being used is the logic model (sometimes called program logic) approach (Razmgir, et al. 2021). This is based on a series of cause-and-effect relationships. The starting point is often a statement of an issue that needs to be addressed. This then leads to the available inputs that could be used, followed by a series of activities. These activities produce outputs (what is delivered), which in turn deliver outcomes (the resulting changes). Ultimately these outcomes will produce impacts on the originating issue. There are many variations on this basic pattern, including categorisation of outcomes as short, middle, or long term. The program logic approach is often used for project planning, but can also be used as an evaluation framework[5], and one which can point to stages of evaluation and opportunities for thinking about sustainability as a KI evolves.

# Conclusion

The process of asking questions and listening for answers regarding the sustainability of KIs is an invitation to managers of emerging and established KIs to be intentional about the persistence of their infrastructures as organizations. Caring deliberately or even strategically about the sustainability of a KI reveals that the only one-size-fits-all solution applicable to every KI revolves around an iterative alignment between infrastructural work and institutional work specific to each KI. As such, demonstrating the long-term value and relevance of a KI to policymakers or funders cannot be reduced to a demonstration of financial compliance and is inseparable from the meaning of sustainability for the key stakeholders as well as the community supporting and using the infrastructure on a daily basis.

Seeking this iterative alignment is also an invitation to KI managers to question the historical preference for the monumental in the funding of KIs by government agencies, encouraging them to embrace the matter of size in terms of relevance rather than expansion for its own sake. Counterintuitively, embracing degrowth may help a KI achieve greater impact, both academically and non-academically, by focusing energy on essential elements—both infrastructural, such as those highlighted by the GORC framework, which identifies interoperability as a crucial lever, and institutional, such as those indicated by the POSI framework, where open governance can unlock the potential of the infrastructure. Finally, asking questions and listening for answers also helps us understand why KI sustainability remains a topic of interest: because forging an institution through individual and collective effort that will withstand the test of time is a target as fluid as time itself.

# Acknowledgements

We would like to thank all participants at the Lorentz Center Workshop "How to Evaluate Emerging Knowledge Infrastructures." for the stimulating conversations which motivated this piece. In particular, we thank our discussion group focusing on sustainability issues, which in addition to the authors included [name removed for review].

We also thank the Lorentz Center itself for providing funding and resources for the workshop. The work of KG on this paper was funded by a Veni grant from the Dutch Research Council (NWO) under the grant number VI.Veni.231S.006. The work of MM on this paper is supported by the NSF National Center for Atmospheric Research, which is a major facility sponsored by the National Science Foundation under Cooperative Agreement No. 1852977.

5An example of how a large, government-funded organization applies this approach is provided by CSIRO, the Australian national science agency. https://www.csiro.au/en/about/corporate-governance/ensuring-our-impact/a-csiro-wide-approach-to-impact

The work of AT on this paper is supported by the Australian Research Data Commons which is enabled by the National Collaborative Research Infrastructure Strategy.

The authors contributed to this article as follows: Conceptualization: KG, KS, JZ, MCV, AT, LB; Funding acquisition: KG, MM; Investigation: KG, JZ; Methodology KG, KS, JZ; Project administration: KG; Writing – original draft: KG, KS, JS, MM, AT; Writing – review & editing: KG, KS, JZ, MCV, MM, AT, LB